\documentclass[aps,pre,reprint]{revtex4-2}
\usepackage{amsmath,amssymb,graphicx,subfigure,xcolor}
\usepackage[english]{babel}
\usepackage[utf8]{inputenc}
\usepackage{epstopdf}

\newcommand{\te}{\theta}

\newcommand{\dte}{\dot{\theta}}
\newcommand{\bte}{\bar{\theta}}
\newcommand{\tte}{\tilde{\theta}}

\newcommand{\tR}{\tilde{R}}
\newcommand{\tP}{\tilde{P}}
\newcommand{\J}{J_{jk}}

\newcommand{\pa}{\partial}

\newcommand{\lan}{\langle}
\newcommand{\ran}{\rangle}

\newcommand{\om}{\omega}

\newcommand{\ph}{\varphi}

\newcommand{\spa}[1]{\qquad\mathrm{#1}\qquad}
\newcommand{\ub}[1]{\underbrace{#1}}

\newcommand{\erf}{\mathrm{erf}}
\newcommand{\cO}{{\mathcal{O}}}

\begin{document}
	
\title{Role of Coupling Asymmetry in the Fully Disordered Kuramoto Model}
\author{Axel Prüser}
\email{axel.prueser@uol.de}
\author{Andreas Engel}
\affiliation{Carl von Ossietzky University Oldenburg, Institut für Physik, D26111 Oldenburg, Germany}

\begin{abstract}
	We investigate the dynamics of phase oscillators in the fully disordered Kuramoto model with couplings of defined asymmetry. The mean-field dynamics is reduced to a self-consistent stochastic single-oscillator problem which we analyze perturbatively and by numerical simulations. We elucidate the influence of the asymmetry on the correlation and response function of the system as well as on the distribution of the order parameter. The so-called volcano transition is shown to be robust with respect to a small degree of coupling asymmetry but to disappear when the antisymmetry in the couplings outweighs the symmetry.
	
\end{abstract}

\maketitle

\section{Introduction}
Networks of coupled phase oscillators exhibit a broad variety of interesting dynamical behaviour~\cite{PiRoKu,Strobook,PiRo}. The homogeneous case of uniform couplings is renowned for the onset of  synchronization at a certain threshold of the interaction strength~\cite{Kubook}. Many aspects and applications of this transition have been investigated over the years~\cite{Stro00,Acebron,PiRo}. Disordered networks in which the  individual oscillators interact via random couplings show even richer dynamics. Depending on statistics, strength, and symmetry of the interactions chaotic behaviour~\cite{SoCrSo88,StRa98}, partial and antiphase synchronization, as well as traveling-wave states may be observed~\cite{HoStro11a,HoStro11b,KloLiStro14,OtStro18,PB}. 

A particularly intriguing question for these random networks concerns the possible existence of an oscillator glass with synchronization into disordered phase patterns~\cite{Daido92,OtStro18,PaGa23,PRE}. In this connection also the relevance of the so-called "volcano transition" found by Daido~\cite{Daido92} on the basis of his pioneering numerical simulations of the Kuramoto model~\cite{Kubook} with random couplings remained an open question~\cite{OtStro18,PaGa23,PRE}. 

In the Kuramoto model each oscillator interacts with every other one. As usual for such a mean-field model the dynamics of the interacting system of many oscillators may be reduced to a self-consistent single oscillator problem. For the case of symmetric but otherwise independent random interactions drawn from a Gaussian distribution this reduction was recently accomplished with the help of the dynamical cavity method borrowed from the theory of spin glasses~\cite{PRE}. The volcano transition was established and its origin clarified. Moreover, arguments were given why its connection to the possible existence of an oscillator glass phase may be rather loose. 

The case of symmetric couplings is somewhat special. In the absence of external driving, i.e., if all natural frequencies of the oscillators coincide, the dynamics for symmetric interactions derive from a Ljapunov function. If, on the other hand, the distribution of couplings allows for a sufficient degree of asymmetry the time evolution becomes chaotic. Moreover, an asymmetry in the couplings is likely to destabilize a supposed glass transition~\cite{CS, SpKi, EiOp94}. Little is known at present about how the volcano transition is modified by asymmetric couplings.

In the present paper we generalize the dynamical cavity approach to a Kuramoto model with random interactions of defined asymmetry. We derive the corresponding self-consistent single-oscillator equation and analyze it perturbatively and by numerical simulations. On the basis of our findings we discuss the modifications of the correlation and response function induced by the coupling asymmetry and analyze the order parameter distribution to investigate the influence of asymmetric couplings on the volcano transition. 

The reduction to a self-consistent single-oscillator problem may also be accomplished by using dynamic generating functionals \cite{MSR,DeDo78} and is often referred to as dynamic mean-field theory \cite{Leticia,SoCr18}. An approach building on the Ott-Antonson ansatz \cite{OA}, on the other hand, fails for the case of full disorder because a diverging number of order parameters were necessary \cite{OtStro18,PaGa23}.

The paper is organized as follows. In section II we define the model and fix the appropriate distribution of couplings. Section III describes the reduction to an effective single-oscillator problem in the form of a dynamically self-consistent stochastic equation of motion. In sections IV and V we give some details about the simulations and the perturbation theory in the coupling strength, respectively, together with a discussion of our results for the correlation and response function. Section VI is devoted to the modification of the order parameter distribution due to the coupling asymmetry and its influence on the volcano transition. Finally, section VII contains our conclusions. 


\section{Model}
The model is defined by the equations
\begin{equation}\label{basiceq}
	\pa_t\te_j(t)=\om_j+h_j(t)+\sum_k \J \sin\big(\te_k(t)-\te_j(t)\big)
\end{equation} 
for the time evolution of phases $\te_j$ of $N$ oscillators $j= 1,...,N$.  The local fields $h_j(t)$ are only introduced for the definition of the susceptibilities,
\begin{equation}\label{defchiN}
    \chi_{jk}(t,t'):=\frac{\delta \te_j(t)}{\delta h_k(t')},
\end{equation}
and will be put equal to zero otherwise. The frequencies $\om_j$ and the initial conditions $\te_j(0)=\te_j^{(0)}$ are drawn independently at random from the respective distributions
\begin{align}\label{Pom}
	P\big(\om\big)&= \frac{1}{\sqrt{2\pi}}\,\exp\left(-\frac{\om^2}{2}\right),\\\label{PJ}
	P\big(\te^{(0)}\big)&=\frac{1}{2\pi},\qquad -\pi\leq\te^{(0)}< \pi\quad .
\end{align}
The couplings are given by
\begin{equation}
\J=\sqrt{\frac{1+\eta}{2}}\J^s+\sqrt{\frac{1-\eta}{2}}\J^a.
\end{equation}
For $j<k$ both $\J^s$ and  $\J^a$ are independent identically distributed Gaussian variables with zero mean and variance $J^2/N$ whereas the values of $\J$ for $j>k$ derive from $\J^s=J_{kj}^s$ and  $\J^a=-J_{kj}^a$. Accordingly, the statistics of the couplings is specified by
\begin{equation}\label{StatJ}
\langle \J \rangle=0, \qquad \langle \J^2 \rangle=\frac{J^2}{N}, \qquad \langle \J J_{kj} \rangle=\eta \frac{J^2}{N}.
\end{equation}
The parameter $\eta\in[-1,1]$ characterizes the degree of asymmetry in the couplings. Particularly, $\eta=1$ stands for the symmetric case, $\J=J_{kj}$, and $\eta=-1$ for antisymmetric couplings, $\J=-J_{kj}$. The value $\eta=0$ denotes the completely asymmetric case with no correlations between $\J$ and $J_{kj}$.

Our interest lies in the exploration of the system's long-time dynamics in the thermodynamic limit $N\to\infty$. Hence the overall coupling strength $J$ and the asymmetry value $\eta$ are the main parameters in the problem.


\section{Reduction to a single-oscillator problem}

Since each oscillator interacts with every other one in a statistically identical way, the model is of mean-field type and may be reduced to an effective, self-consistent single-oscillator problem. To achieve this goal, we apply the dynamical cavity method \cite{MPV,PRE}.

We consider system~\eqref{basiceq} for one particular realization of frequencies, initial conditions and couplings and add one {\em new} oscillator, $j=0$, with phase $\te_0(t)$ and new quenched random variables $\om_0,\te_0^{(0)}, J_{0k}$, and $J_{k0}$. The presence of the additional oscillator triggers slight perturbations
\begin{equation*}
    \te_j\to\tte_j=\te_j+\delta\te_j
\end{equation*}
in the dynamics of the pre-existing ones. Given that $J_{k0}=\cO(1/\sqrt{N})$, these perturbations are small and may be treated using linear response, 
\begin{equation}\label{defdelte}
	\delta\te_j(t)=\sum_l\!\int_0^t\!\! dt'\, 
	\chi_{jl}(t,t') J_{l0}\sin\big(\te_0(t')-\te_l(t')\big).
\end{equation}
To leading order in $N$ the equation governing the motion of the new oscillator then reads
\begin{widetext}
	\begin{equation}\label{eqte0a}
		\pa_t\te_0(t)= \om_0+h_0(t)+\sum_k J_{0k} \sin\big(\te_k(t)-\te_0(t)\big)
		+\sum_{k,l} J_{0k} J_{l0}\int_0^t\!\! dt'\,\chi_{kl}(t,t')\cos\big(\te_k(t)-\te_0(t)\big)\sin\big(\te_0(t')-\te_l(t')\big).
	\end{equation}
\end{widetext}
The third and the fourth term of this equation are random functions of time. Due to the statistical independence of the new couplings $J_{0k},\,J_{k0}$ and the old phases $\te_j,\, j=1,\dots, N$ both are large sums of uncorrelated contributions. We will therefore assume that in the limit $N\to\infty$ they are Gaussian random functions with statistical properties fixed by their averages and correlations. 
 
Since the third term involves just one coupling $J_{0k}$ it is independent of $\eta$. Hence we find as in the case $\eta=1$  \cite{PRE}
\begin{align*}
	\sum_k J_{0k} \sin&\big(\te_k(t)-\te_0(t)\big)\\
	=&J\cos\te_0(t)\,\xi_2(t)-J\sin\te_0(t)\,\xi_1(t),
\end{align*}
where $\xi_1$ and $\xi_2$ are two independent Gaussian noise sources. Denoting by $\lan\dots\ran$ the combined average over {\em all} $\J$, (i.e., $j,k=0,\dots, N$) and the {\em old} $\om_j,\te_j^{(0)}$, (i.e., $j=1,\dots, N$), their average and correlations are given by
\begin{equation*}
	\lan\xi_a(t)\ran\equiv 0,\qquad \lan\xi_a(t)\xi_b(t')\ran=\delta_{ab}\,C(t,t'),
\end{equation*}
with $a,b=1,2$ and 
\begin{align}\label{defCprel}
	C(t,t')&:=\frac{1}{2N}\sum_j\lan\cos \bte_j(t,t')\ran,
\end{align} 
where the abbreviation $\bte_j(t,t'):=\te_j(t)-\te_j(t')$ was introduced.

The fourth term in Eq.~\eqref{eqte0a} involves the product $J_{0k} J_{l0}$ which is sensitive to the asymmetry parameter $\eta$. Its statistical properties are analyzed in Appendix \ref{AA} where it is shown that its fluctuations are negligible for $N\to\infty$ such that it may be replaced by its average. Plugging these results into~\eqref{eqte0a} the differential equation for the new oscillator assumes the form
\begin{widetext}
\begin{equation}\label{LEsc1}
	\pa_t\te_0(t)=\om_0+h_0(t)+J \cos\te_0(t)\,\xi_2(t)-J\sin\te_0(t)\,\xi_1(t)
	-\eta J^2\int_0^t\!\! dt'\,R(t,t')\sin\bte_0(t,t'),
\end{equation} 
\end{widetext}
with, cf.~\eqref{resR2},
\begin{equation}\label{defR2}
    R(t,t')=\frac{1}{2N}\sum_k \lan\chi_{kk}(t,t') \cos\bte_k(t,t')\ran.
\end{equation}

Hence, the time evolution of the new phase $\te_0(t)$ is governed by a Langevin equation, with correlation and response function determined by the dynamics of the original $N$-oscillator system in the \textit{absence} of $\te_0$. To close the  argument, we observe that in the averaged $(N+1)$-oscillator system $\te_0$ is in no way special. Consequently, the averages in Eqs.~\eqref{defCprel} and~\eqref{defR2} can equivalently be taken over the dynamics of $\te_0(t)$ itself. The  self-consistent single-oscillator problem is therefore given by equation~\eqref{LEsc1}, together with
\begin{align}\label{defC}
		\lan\xi_a(t)\ran&\equiv 0,\qquad \lan\xi_a(t)\xi_b(t')\ran=\delta_{ab}\,C(t,t'),\\\label{defC1}
	C(t,t')&:=\frac{1}{2}\lan\cos \bte_0(t,t') \ran,\\\label{defR}
	R(t,t')&:=\frac{1}{2}\lan\chi_{00}(t,t')\cos  \bte_0(t,t') \ran,\\\nonumber
	\chi_{00}(t,t')&:=\frac{\delta\te_0(t)}{\delta h_0(t')}.
\end{align} 
 The average $\lan\dots\ran$ is now over the realizations of the $\xi_a$, $\om_0$ and $\te_0^{(0)}$. For large values of $t$ and $t'$, the functions $C, R$, and $\chi$ will only depend on $\tau:=t-t'$.
The equation of motion for $\chi_{00}(t,t')$ is
\vspace{-3mm}
\begin{widetext}
	\begin{align}\nonumber
		\pa_t\chi_{00}(t,t')=\frac{\delta\pa_t\te_0(t)}{\delta h_0(t')}\Big|_{h(t)\equiv 0}
		=\delta(t-t')-J\chi_{00}(t,t')&\big(\sin\te_0(t)\,\xi_2(t)+\cos\te_0(t)\,\xi_1(t)\big)\\\label{eomchi}
		&-\eta J^2\int_0^t\!\!dt''\,R(t,t'')\cos\bte_0(t,t'')\big(\chi_{00}(t,t')-\chi_{00}(t'',t')\big)
	\end{align} 
\end{widetext}
with initial condition $\chi_{00}(t,t')=0\spa{for} t<t'$. This equation has to be solved in parallel with~\eqref{LEsc1} in order to obtain the response function $R(t,t')$ according to Eq.~(\ref{defR}). 

This completes the reduction of the original system~\eqref{basiceq} to a self-consistent dynamical single-oscillator problem. The index $0$ will be omitted from now on. Despite the reduction achieved, Eqs.~\eqref{LEsc1},\eqref{defC}-\eqref{eomchi} still represent a complicated system of equations. To further analyze it we employ numerical simulations as described in the next section and perturbation theory in the couplings strength $J$ which is detailed in section~\ref{sec:pt}.


\section{Numerical simulations}\label{sec:num}

\begin{figure*}[t]
	\centering
	\subfigure{\includegraphics[scale=0.44]{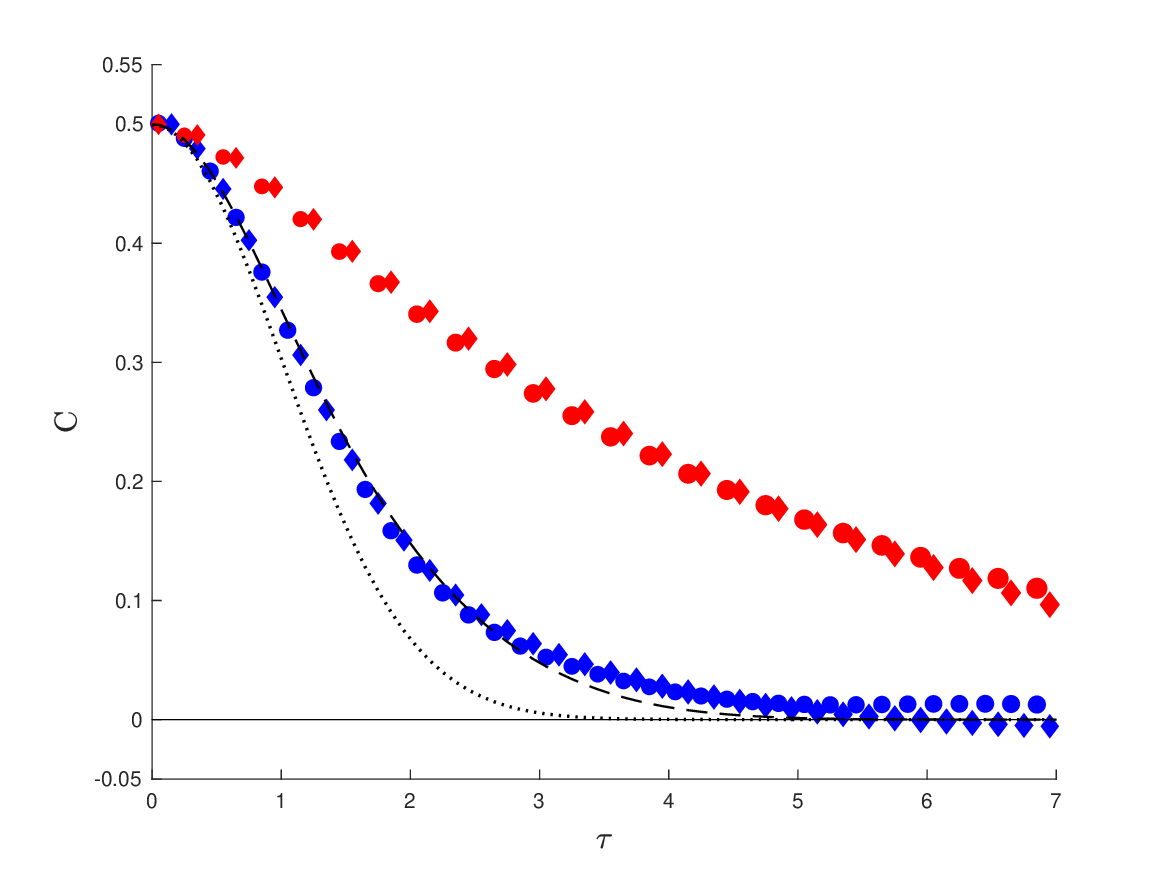}}\quad
	\subfigure{\includegraphics[scale=0.44]{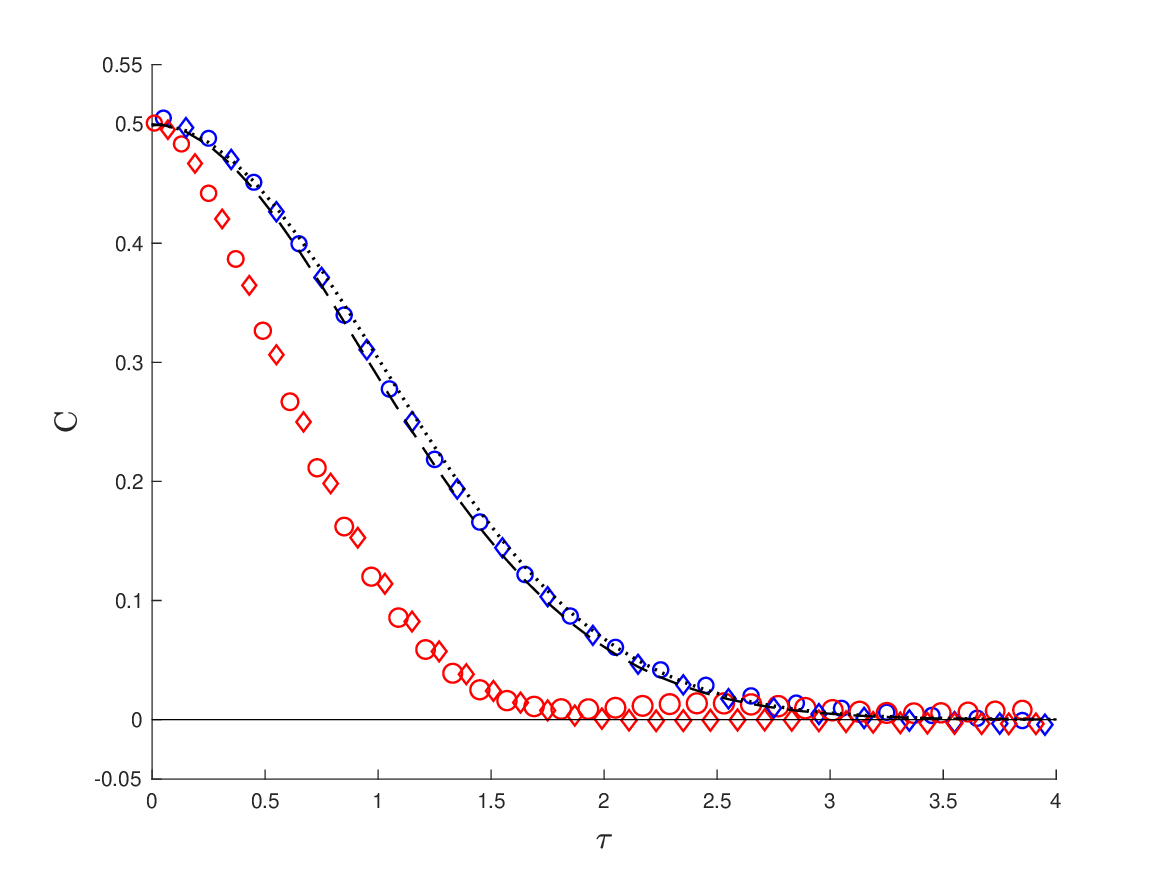}}
	\caption{Correlation function $C(\tau)$ of the self-consistent noise sources $\xi_1(t)$ and $\xi_2(t)$ in~\eqref{LEsc1} for $\eta=0.5$ (left) and $\eta=-0.5$ (right) for $J=.8$ (blue symbols) and $J=4$ (red symbols), respectively. Circles show data obtained from simulations of the single-oscillator dynamics with up to $3.7\cdot 10^4$ trajectories and using~\eqref{defC1}, diamonds represent results for an $N=10^3$ oscillator network by employing~\eqref{defCprel}, averaged over 37 disorder realizations. Statistical errors are of symbol size or smaller. Also displayed is the lowest order perturbation result $C_0(\tau)$ given by~\eqref{resC0} (dotted line) and, for $J=0.8$, the second order result~\eqref{expC} (dashed line).} 
	\label{fig1}
\end{figure*}

The obtained self-consistent single-oscillator problem is amenable to numerical simulations. The method was originally proposed by Eißfeller and Opper~\cite{EiOp94} for spin systems and later extended to continuous degrees of freedom by Roy~et~al.~\cite{Roy}. One starts with initial guesses for $C(t,t')$ and $R(t,t')$, generates a family of noise trajectories 
$\{\xi_1(t)\}$ and $\{\xi_2(t)\}$ according to~\eqref{defC} and using these determines an ensemble of trajectories $\{\te(t)\}$ by solving~\eqref{LEsc1} and~\eqref{eomchi}. From these one determines refined approximations for $C(t,t')$ and $R(t,t')$ via~\eqref{defC1} and ~\eqref{defR} and iterates the procedure until convergence is reached. More details can be found in \cite{PRE}. Note that this kind of numerical simulations is rather different from the direct numerical solution of the original system~\eqref{basiceq} for different realizations of the quenched disorder with subsequent averaging as performed by Daido~\cite{Daido92}. In particular, apart from frequency and initial condition of the new oscillator all random parameters of the $N$-oscillator system are subsumed in the noise sources $\xi_1(t)$ and $\xi_2(t)$. Moreover, no finite-size effects come into play~\cite{EiOp94} because the limit $N\to\infty$ was already explicitly taken in the derivation of~\eqref{LEsc1}.

We performed simulations of this kind by implementing a basic Euler scheme with stepsize $\Delta t=0.05$ to solve Eqs.~\eqref{LEsc1} and~\eqref{eomchi}. For small $\eta$ and large $J$ the stepsize was reduced to $\Delta t=0.01$. Simulations with up to $800$ time steps were used to determine the self-consistent forms of $C$ and $R$. Once these were found, much longer trajectories $\te(t)$ with $5\cdot 10^5$ time steps were generated to collect the data shown in the figures.

To validate our code as well as the assumptions made in the derivation of~\eqref{LEsc1} we first compare the correlation function $C(\tau)$ of the single-oscillator dynamics as obtained from~\eqref{defC1} with the one resulting from a simulation of the original $N$-oscillator network and using~\eqref{defCprel}. The numerical solution of the  system~\eqref{basiceq} was accomplished using a fourth-order Runge-Kutta method. Step size and trajectory lengths are the same as in the single-oscillator problem.

Fig.~\ref{fig1} shows the results obtained for the correlation function for $\eta=.5$ (left) and $\eta=-.5$ (right), in both cases for two different values of the interaction strength $J$. The agreement between $N$- and single-oscillator simulations is in all four cases quite satisfactory. For positive $\eta$ long-time correlations build up in the noise-sources $\xi_1(t)$ and $\xi_2(t)$ when $J$ is increased reflecting the tendency of the $N$-oscillator system to phase synchronization. This is qualitatively similar to the symmetric case, $\eta=1$, as discussed in~\cite{PRE}, but quantitatively less pronounced. On the contrary, for predominantly negative coupling correlations, as described by $\eta=-.5$, the correlation time decreases with increasing $J$, in accordance with a growing chaoticity of the dynamics \cite{StRa98}.

\begin{figure}[t]
	\includegraphics[width=.44\textwidth]{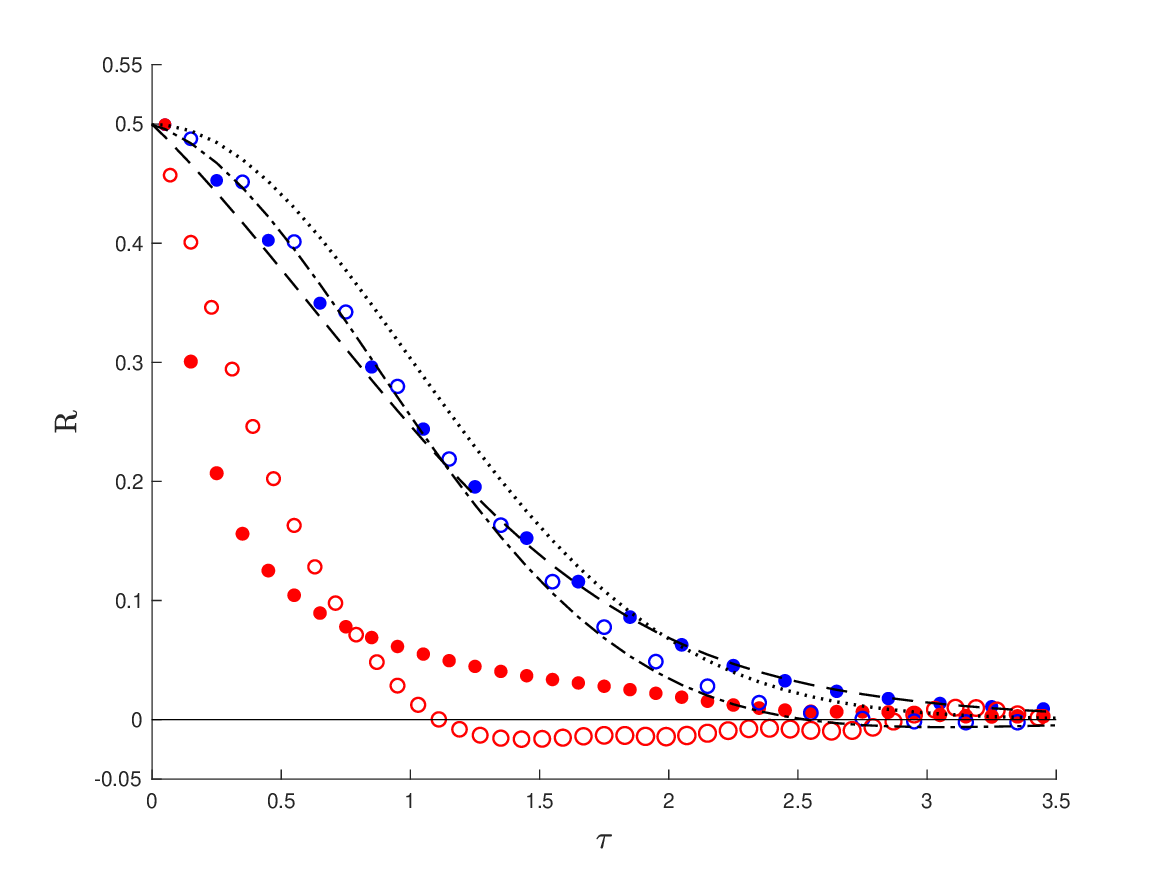}
	\caption{Response function $R(\tau)$ appearing in Eqs.~\eqref{LEsc1} and \eqref{eomchi} for $\eta=0.5$ (filled circles) and $\eta=-0.5$ (empty circles). Blue symbols are for $J=0.8$ and red ones for $J=4$. Shown are results from simulations of the single-oscillator dynamics and Eq.~\eqref{defR} using up to $3.7\cdot 10^4$ trajectories. Statistical errors are of symbol size or smaller. Included are also the lowest order perturbation result $R_0(\tau)$ (dotted line) and, for $J=0.8$, the second order results~\eqref{expR} for $\eta=0.5$ (dashed line) and $\eta=-0.5$ (dash-dotted line).}
	\label{fig2}
\end{figure}

The different forms of the response function $R(\tau)$ for the same combination of $\eta$ and $J$ values as in Fig.~\ref{fig1} are shown in Fig.~\ref{fig2}. Only simulations of the single-oscillator dynamics were used and no comparison with the $N$-oscillator system was attempted in this case. For both  $\eta=.5$ and $\eta=-.5$ the response function falls off more rapidly for small values of $\tau$ when $J$ is increased. For positive $\eta$ this tendency is somewhat counterbalanced by the build-up of a long-time tail as characteristic also for the symmetric case $\eta=1$. For negative $\eta$ no such tail is observed and the response function decays to zero rather quickly. This behaviour is corroborated by our pertubative results to be discussed next.

\section{Perturbation Theory}\label{sec:pt}
To investigate the single-oscillator problem with the help of perturbation theory in the coupling strength $J$ we use the expansion
\begin{equation}\label{pertte}
	\te(t)=\te_0(t)+J\te_1(t)+J^2\te_2(t)+\cO(J^3).
\end{equation} 
as well as
\begin{align}\label{expC}
	C(t,t')&=C_0(t,t')+JC_1(t,t')+J^2C_2(t,t')+\cO(J^3),\\\label{expR}
	R(t,t')&=R_0(t,t')+J R_1(t,t')+J^2 R_2(t,t')+\cO(J^3)\\\nonumber
 \chi(t,t')&=\chi_0(t,t')+J \chi_1(t,t')+J^2 \chi_2(t,t')+\cO(J^3).
\end{align}
Note that the index now denotes the order of perturbation theory. It is also convenient to define the Gaussian random functions
\begin{align}\label{defphpsi}
	\ph(t)&:=\cos\te_0(t)\,\xi_2(t)-\sin\te_0(t)\,\xi_1(t),\\\nonumber
	\psi(t)&:=-\sin\te_0(t)\,\xi_2(t)-\cos\te_0(t)\,\xi_1(t).
\end{align}
Their statistical properties derive from~\eqref{defC}:
\begin{align*}\nonumber
	\lan\ph(t)\ran_\xi&\equiv\lan\psi(t)\ran_\xi\equiv 0,\\
	\lan \ph(t)\ph(t')\ran_\xi&=\lan \psi(t)\psi(t')\ran_\xi=C(t,t')\cos\bte_0(t,t'),\\
	\lan \ph(t)\psi(t')\ran_\xi&=-\lan \psi(t)\ph(t')\ran_\xi=C(t,t')\sin\bte_0(t,t').
\end{align*} 
Plugging the perturbation expansions into~\eqref{LEsc1} and collecting powers of $J$ we find
\begin{align}\label{eq00}
	\dte_0(t) &= \om+h(t)\\\label{eq01}
	\dte_1(t) &= \ph(t)\\\label{eq02}
	\dte_2(t) &= \psi(t)\te_1(t)-\eta \int_0^t\!\!dt'\, R_0(t,t')\sin\bte_0(t,t').
\end{align}
From \eqref{eq00} we immediately get the lowest order result
\begin{equation}\label{res0}
	\te_0(t)=\om t +\int_0^t\!\!dt'\, h(t') + \theta^{(0)}
\end{equation} 
and therefore
\begin{equation*}
  \chi_0(t,t')=\frac{\delta \te_0(t)}{\delta h(t')}=\kappa(t-t'):=
    \begin{cases}
		1 &  t>t'\\
		0 & \mathrm{else}
	\end{cases}\; ,
\end{equation*}
where the Heaviside-function was denoted by $\kappa$. Setting now $h(t)\equiv 0$ and $\tau=t-t'$, and using~\eqref{Pom}, we find
\begin{equation}\label{resC0}
	C_0(\tau)=\frac{1}{2}\lan\cos\om\tau\ran_{\om,\te^{(0)}}=\frac{1}{2}\,e^{-\frac{\tau^2}{2}}
\end{equation}
as well as 
\begin{equation}\label{resR0}
    R_0(\tau)=\frac{1}{2}\lan\chi_0\cos\om\tau\ran_{\om,\te^{(0)}}=\kappa(\tau)\,C_0(\tau).
\end{equation}
The first order terms, $C_1$ and $R_1$, each involve a single noise term $\xi_a$, cf.~\eqref{eq01} and~\eqref{defphpsi}, and therefore average to zero. Hence to zeroth and first order in $J$ we obtain the same results for $\te(t), C(\tau)$ and $R(\tau)$ as in the case $\eta=1$, cf.~\cite{PRE}. This was to be expected since the first appearance of $\eta$ is in the equation of motion~\eqref{eq02} for $\te_2(t)$.

To second order we have 
\begin{equation}\label{C2}
	C_2 = -\frac{1}{2}\Big[\lan\sin\bte_0\,\bte_2\ran+\frac{1}{2}\lan\cos\bte_0\,\bte_1^2\ran\Big],
\end{equation} 
and 
\begin{equation}\label{res0R2} 
	R_2(\tau)=\kappa(\tau) C_2(\tau)+\frac{1}{2}\lan \chi_2\,\cos\bte_0\ran,
\end{equation} 
where the averages over the $\xi_a$ are with correlations described by $C_0$. The term involving $\chi_1$ in $R_2$ drops out. The remaining averages are calculated in Appendix~\ref{AB} and give rise to 
\begin{widetext}
	\begin{equation}\label{resC2erf}
	C_2(\tau)=\frac{\tau\sqrt{\pi}}{32}\,e^{-\frac{1}{4}\tau^2}\Big[
	(4\eta+5)\,\erf\left(\frac{\tau}{2}\right)-3 \, \erf\left(\frac{3\tau}{2}\right)\Big] 
	+ \frac{1}{16}\,e^{-\frac{1}{2}\tau^2}\Big[1-e^{-2\tau^2}\Big]
\end{equation} 
and 
\begin{equation}\label{resR2erf}
	R_2(\tau)=\kappa(\tau)\left(\frac{\tau\sqrt{\pi}}{32}\,e^{-\frac{1}{4}\tau^2}
	\Big[(3\eta +4)\,\erf\left(\frac{\tau}{2}\right) + 3\eta\,\erf\left(\frac{3\tau}{2}\right)-4(\eta+1)\Big] - 
	\frac{\eta}{16}\,e^{-\frac{1}{2}\tau^2}\Big[1-e^{-2\tau^2}\Big]\right).
\end{equation} 
\end{widetext}

For $J=0.8$ the perturbative results for $C(\tau)$ and $R(\tau)$ up to second order have been added to Figs.~\ref{fig1} and~\ref{fig2}. For both correlation and response function  the agreement with the simulation results is very good. This gives further credit to our numerical treatment of the single-oscillator problem. Unfortunately, second order perturbation theory is not sufficient to get sensible results up to $J=4$ so that no comparison with the simulations at this larger value of the coupling strength is possible.  

Nevertheless, we can use the expressions for $C_2$ and $R_2$ to elucidate the differences in the dynamics for positive and negative values of $\eta$ for small $J$. From~\eqref{resC2erf} we infer that $C_2(\tau)$ decreases for all $\tau$ when $\eta$ decreases from one. This is again in accordance with the growing degree of chaoticity in the dynamics. $R_2$, on the other hand, exhibits an increase for small $\tau$ when $\eta$ gets smaller as described by the last term in~\eqref{resR2erf}. For larger values of $\tau$, however, the first term in this expression dominates due to the slower exponential decrease and therefore also $R_2(\tau)$ decreases with decreasing $\eta$. This behaviour is consistent with our numerical results, cf. Fig.~\ref{fig2}. The simulations also suggest that this qualitative trend holds for larger values of $J$ as well: asymptotically the response function for $\eta=-0.5$ decays faster than the one for $\eta=0.5$.

Notably, the case $\eta=-1$ of completely antisymmetric couplings gives rise to 
\begin{equation}\label{CRanti}
    R_2(\tau)=\kappa(\tau) C_2(\tau),
\end{equation}
such that the simple relation~\eqref{resR0} between correlation and response function characteristic for the trivial case $J=0$ seems to hold true also in the realm of interacting oscillators as long as $\eta=-1$. This is confirmed also by our numerical results for larger values of $J$. It would therefore be interesting to see whether~\eqref{CRanti} holds to all orders in perturbation theory.
 

\section{Order parameter}

To characterize the system's ability to synchronize in disordered patterns, Daido introduced the complex local fields \cite{Daido92}
\begin{equation}\label{defpsi}
	\Psi_j(t):=r_j(t)\,e^{i\phi_j(t)}:=\frac{1}{J}\sum_k\J\,e^{i\te_k(t)}.
\end{equation}
For constant couplings $J_{jk}$ this expression reduces to the standard order parameter $\Psi$ of the original Kuramoto model~\cite{Kubook}. For random interactions the coupling term $\J$ under the sum is crucial. The fields $\Psi_j(t)$ are different for different oscillators and hence their distribution will serve as order parameter. In this respect the role of modulus and phase are different: whereas the distribution $P(r)$ of the moduli $r_j$ describes the general tendency of the system to synchronize into disordered phase patterns, the specific pattern arising is characterized by the distribution of phases $\phi_j$.  

\begin{figure}[t]
	\includegraphics[width=.44\textwidth]{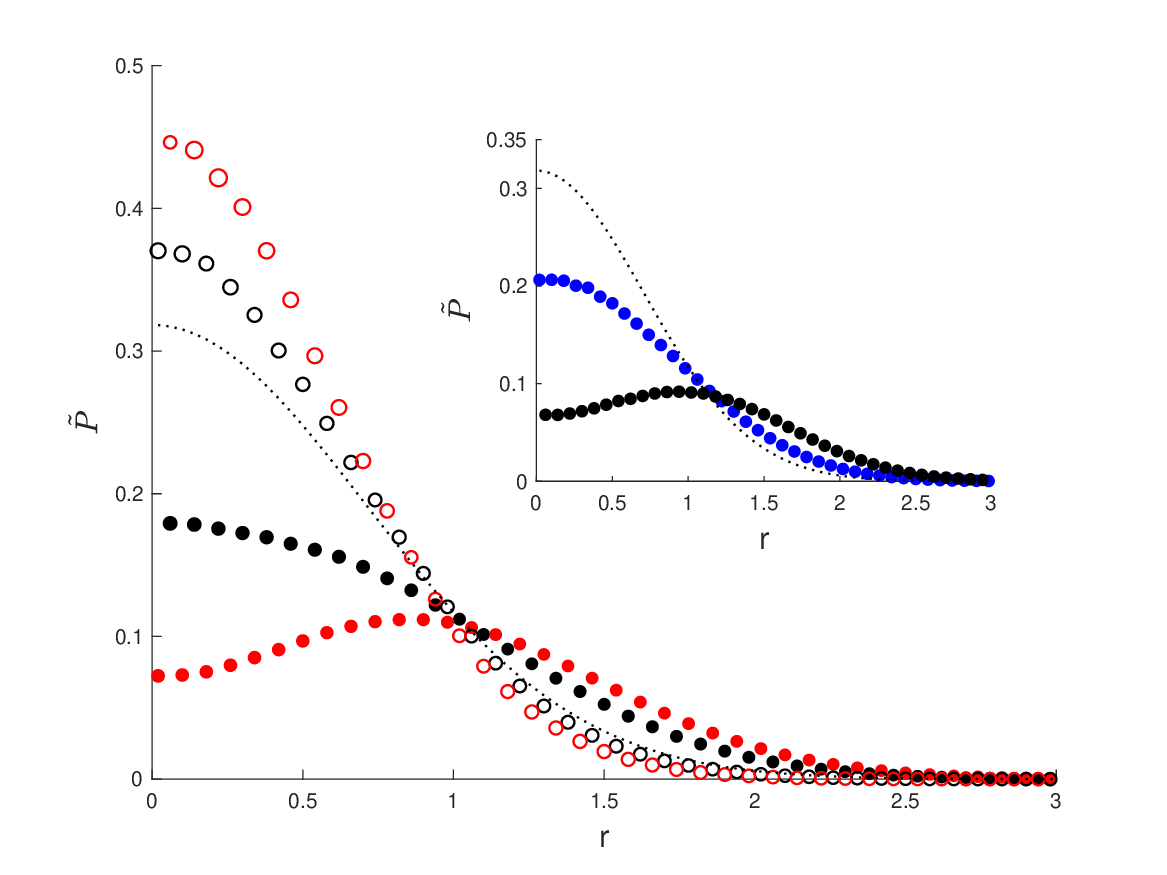}
	\caption{Distribution $\tP(r)$ of local field moduli as defined in~\eqref{tP} comparing the cases $\eta=0.5$ (filled circles) and $\eta=-0.5$ (empty circles) for $J=1.75$ (black) and $J=4$ (red). The inset shows the behaviour for $\eta=1$ and $J=.8$ (blue) and  $J=1.75$ (black). The black dotted line in both figures represent $\tP_0(r)$ as given by~\eqref{tP0} which characterizes the trivial situation $J=0$. Data are from numerical simulations of up to $5.8\cdot 10^4$ trajectories of the single-oscillator problem. Statistical errors are of symbol size or smaller.}
    \label{fig3}
\end{figure}

In his numerical analysis, Daido focused on the modified distribution of field moduli
\begin{equation}\label{tP}
	\tP(r):=\frac{P(r)}{2\pi r},
\end{equation}
because it shows the so-called volcano transition - a topological change of $\tP(r)$ with the maximum moving from $r=0$ to a value $r>0$ when the coupling strength $J$ increases beyond a critical value. In the inset of Fig.~\ref{fig3} this transition is shown for $\eta=1$.

In the present section, our focus is on the influence of the coupling asymmetry $\eta$ on the distribution $\tP(r)$ of local field moduli in general and on the occurrence of the volcano transition in particular. To this end we consider 
the local field of the newly added oscillator given by
\begin{equation}\label{defP0}
	\Psi(t):=\frac{1}{J}\sum_j J_{0j}\, e^{i\tte_j(t)}.
\end{equation}
Exploiting the smallness of $\delta\te_j$ and using~\eqref{defdelte} as well as the results of appendix~\ref{AA} we find 
\begin{widetext}
\begin{align}\nonumber
	\Psi(t)=&\frac{1}{J}\sum_j  J_{0j}\, e^{i\te_j(t)}\big(1+i\,\delta\te_j(t)+\dots\big)\\\nonumber
	=&\frac{1}{J}\sum_j J_{0j}\, e^{i\te_j(t)}+i\frac{1}{J}\sum_j J_{0j}\, e^{i\te_j(t)}\sum_l\int_0^t\!\! dt'\,\chi_{jl}(t,t') J_{l0}\sin(\te'-\te'_l)\\
	=&\frac{1}{J}\sum_j J_{0j}\, e^{i\te_j(t)}+\eta \,J\int_0^t\!\! dt'\,R(t,t')\,e^{i\te(t')}\\
	=&\xi(t)+\zeta(t).\label{decomppsi}
\end{align}
\end{widetext}
Here $\xi(t):=\xi_1(t)+i\xi_2(t)$, with the same Gaussian noise sources $\xi_1(t)$ and $\xi_2(t)$ that already appeared in~\eqref{LEsc1}. 

Therefore, $\Psi(t)$ is the sum of two complex random variables. The first one, $\xi(t)$, is a complex Gaussian noise with zero mean and correlations
\begin{equation*}
    \lan\xi(t)\xi(t')\ran\equiv0,\qquad \lan\xi(t)\xi^*(t')\ran=2C(t,t').
\end{equation*}
The second one is given by
\begin{equation}\label{defzeta}
    \zeta(t):=\eta \,J\int_0^t\!\! dt'\,R(t,t')\,e^{i\te(t')}.
\end{equation}
Note that there are complicated nonlocal correlations between $\zeta(t)$ and $\xi(t')$ due to the dependence of $\te(t)$ on $\xi(t')$ for all $t'<t$. 

Simulating the single-oscillator dynamics as described in section~\ref{sec:num}, we have compiled histograms of $\tP(r)$ according to~\eqref{decomppsi} and~\eqref{defzeta} for different values of $\eta$ and $J$. The results are shown in Fig~\ref{fig3}. As can be seen, the shape modification of $\tP(r)$ with increasing $J$ is strikingly different for $\eta>0$ and $\eta<0$, respectively. In the former case, $\tP(r)$ decreases for small values of $r$ and increases for larger ones when the coupling gets stronger. This is similar to the behaviour for $\eta=1$ and paves the way for the volcano transition. In fact, this transition is observed in our plots for $\eta=.5$. It occurs at $J_c\simeq 3$, which is larger than the critical value $J_c\simeq 1.3$ for $\eta=1$ \cite{Daido92,PRE} as expected. For other positive values of $\eta$ we find a similar behaviour. These findings are in agreement with the numerical simulations reported in~\cite{PaGa23} (see their figure 6) showing that an asymmetry $\eta\gtrsim .2$ in the couplings typically hinders the volcano transition without suppressing it.  Our value $J_c\simeq 3$ for $\eta=0.5$ agrees well with the one obtained in \cite{PaGa23}. We therefore conclude that the volcano transition persists down to rather small positive values of $\eta$ with the critical value $J_c$  increasing with decreasing $\eta$. 

For negative $\eta$, on the other hand, $\tP(r)$ {\em increases} for small values of $r$ with growing $J$ and {\em decreases} at larger $r$ which renders a volcano transition impossible. This is exemplified by the results for $\eta=-.5$ shown in Fig.~\ref{fig3}. 

\begin{figure}[t]
	\includegraphics[width=.44\textwidth]{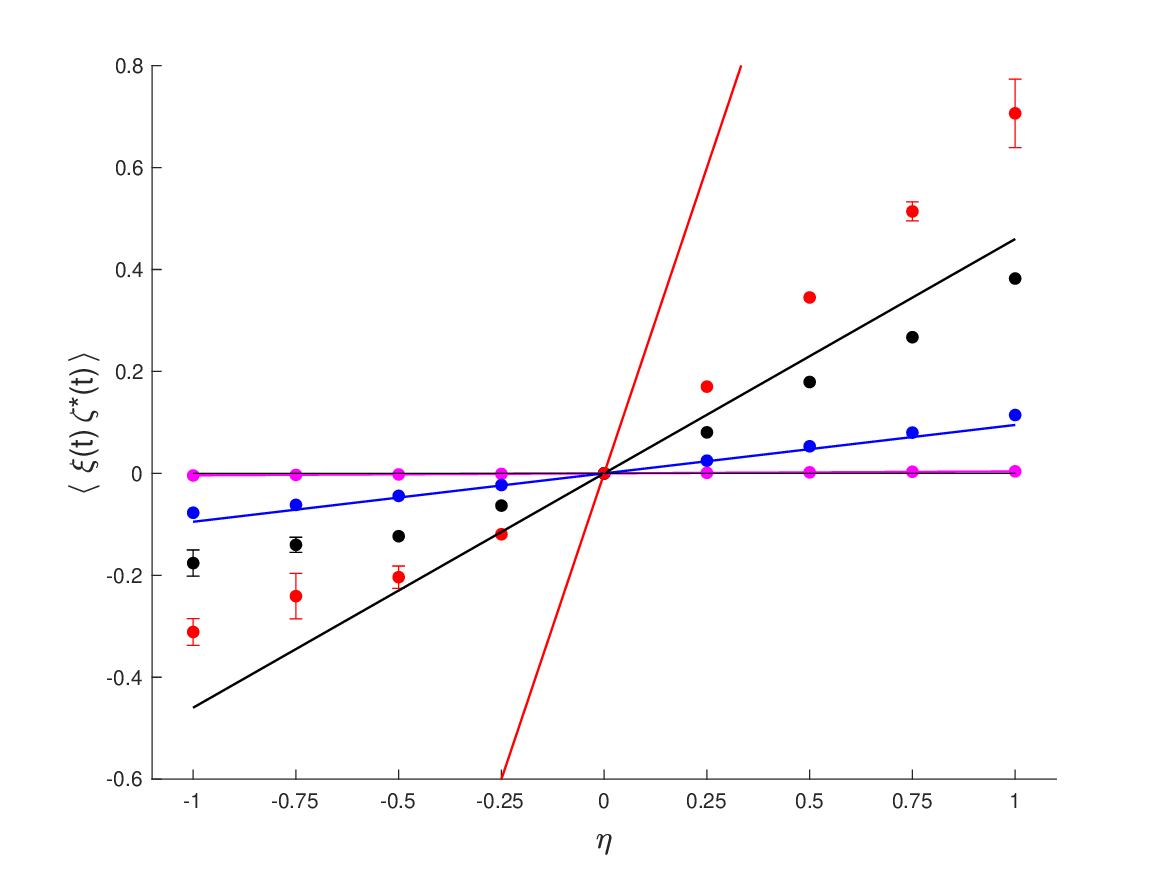}
	\caption{Correlation $\lan \xi(t)\,\zeta^*(t)\ran$ between the two components of the local field as defined in~\eqref{decomppsi} as a function of the coupling asymmetry for $J=0.16$ (magenta), $J=0.8$ (blue), $J=1.75$ (black) and $J=4$ (red). Symbols show results from numerical simulations of the single-oscillator dynamics using up to $5 \cdot 10^3$ trajectories. Statistical errors are given by error bars or the symbol size. The lines depict the approximate  analytical behaviour~\eqref{corr} valid for small $J$.}
	\label{fig4}
\end{figure}

To get some intuition about the behaviour of $\tP(r)$ at different degrees of asymmetry, it is instructive to consider the relative orientation between the two components $\xi(t)$ and $\zeta(t)$ in the decomposition~\eqref{decomppsi}. It is characterized by the correlation $\lan\xi(t)\zeta^*(t)\ran$ which is shown in Fig.~\ref{fig4} as function of $\eta$ for different values of $J$. As is clearly seen, the absolute value of these correlations increases with increasing coupling strength and absolute value of $\eta$ whereas their {\em sign} is the same as that of $\eta$. In appendix~\ref{AC} the approximate relation 
\begin{equation}\label{corr}
    \lan\xi(t)\zeta^*(t)\ran\simeq \, 0.15\, \eta \, J^2,
\end{equation}
valid for small $J$, is derived that corroborates this observation and agrees well with the numerical data up to $J \simeq 0.8$. Now, positive correlations between $\xi$ and $\zeta$ induce a larger spread of $\tP(r)$, connected with a decrease at small values of $r$, eventually giving rise to the volcano transition. That the correlations get weaker with increasing deviation of $\eta$ from one again agrees with the fact that the critical coupling strength for the transition increases. 

For $\eta<0$, on the other hand, the anti-correlation between $\xi$ and $\zeta$ favours small values of $r$, resulting in the observed increase of the maximum of $\tP(r)$ at $r=0$. 

It is finally interesting to consider the special cases $\eta=0$ and $\eta=-1$. As characteristic for completely asymmetric couplings, the response term in~\eqref{LEsc1} is absent for  $\eta=0$~\cite{CS1}. No tendency for synchronization can hence be expected in this case.  Correspondingly, $\zeta$ is identically zero and the local field $\Psi(t)$ is just given by the noise term $\xi(t)$. Its modulus is then Rayleigh distributed and using $C(0)=0.5$ we find  
\begin{equation}\label{tP0}
\tP_0(r)=\frac{1}{\pi}\,e^{-r^2}
\end{equation}
for all values of $J$. 

Surprisingly, this is true also for $\eta=-1$. The reason and its possible connection with the relation $R(\tau)=\kappa(\tau) C(\tau)$ for $\eta=-1$ discussed at the end of section~\ref{sec:pt} remained unclear to us. 


\section{Conclusion}

In this work we have studied the dynamics of $N$ randomly coupled phase oscillators and analyzed the impact of a coupling  asymmetry on the ability of the system to synchronize. To this end we derived a stochastic equation of motion for a single oscillator mimicking a typical member of the $N$-oscillator system in the thermodynamic limit $N\to\infty$. This was done by generalizing the dynamical cavity method of~\cite{PRE} to couplings with definite asymmetry quantified by
\begin{equation*}
	\eta=\frac{\lan J_{jk} J_{kj}\ran}{\lan J^2_{jk}\ran}.
\end{equation*}
The reliability of our method was demonstrated by comparing simulations results of the original $N$-oscillator system and the self-consistent single-oscillator dynamics as well as by the agreement between numerical and approximate analytic results stemming from a perturbation theory.

Quite generally, the correlation and response functions that characterize the dynamics are found to decay faster to zero when $\eta$ decreases from one. This is in accordance with the fact that the dynamics becomes more chaotic, as described by its Lyapunov exponent, when the asymmetry in the couplings increases. Nevertheless, the qualitative behaviour of correlation and response function and also of the order parameter distribution remains similar to the symmetric case for values of $\eta$ down to $\eta\simeq.2$. In particular, the volcano-transition in the distribution $\tP(r)$, defined in~\eqref{tP}, persists, although its onset is shifted to higher values of the coupling strength $J$. These findings are in accordance with the numerical investigations described in~\cite{PaGa23}. Since it is generally believed that already a small deviation of $\eta$ from one precludes a freezing transition~\cite{CS,SpKi,EiOp94}, we come again to the conclusion that there is no fundamental connection between the volcano-transition and the existence of an oscillator glass phase. 

For negative values of $\eta$, i.e., if antisymmetry between the couplings dominates, the situation is quite different. Both correlation and response function become rather short ranged and the volcano transition is absent. Accordingly, there is no tendency to synchronization in the usual sense. Nevertheless, the interaction between the oscillators modifies the dynamics, and the distribution of phases differs from the completely random case $\te_j(t)=\om_jt+\te_j^{(0)}$ arising for $J=0$. The fact that small values of the order parameter modulus $r$ become more frequent with increasing
coupling strength $J$ hints on some overall repulsion between the phases of the different oscillators.

A peculiar case is given by completely antisymmetric couplings, $J_{jk}=-J_{kj}$, corresponding to $\eta=-1$. In this case the order parameter distribution does not depend on the coupling strength $J$, and correlation and response function seem to coincide for positive arguments. It would be interesting to find the intuitive reason for this rather surprising behaviour.

\section*{Acknowledgments}

We would like to thank Sebastian Rosmej and Heiko Rieger for stimulating discussions.


\begin{widetext}
\onecolumngrid

\appendix
\section{}
\label{AA}

We write the fourth term in~\eqref{eqte0a} as $\int_0^t \! dt' R_\mathrm{tot}(t,t')$ where 
\begin{align*}
	R_\mathrm{tot}(t,t'):=\sum_{k,l} J_{0k} J_{l0}\chi_{kl}(t,t') \cos\big(\te_k(t)-\te_0(t)\big)\sin\big(\te_0(t')-\te_l(t')\big).
\end{align*} 
With the abbreviations  $\te_j:=\te_j(t)$, $\te'_j:=\te_j(t')$ and $\bte_j(t,t'):=\te_j-\te'_j$ we find
\begin{align*}
	R_\mathrm{tot}(t,t')=&-\frac{1}{2}\sum_{k,l} J_{0k} J_{l0}\chi_{kl}(t,t') \cos(\te_k-\te'_l)\sin\bte_0
	=:-\eta J^2 R(t,t') \sin\bte_0
\end{align*}
where
\begin{align*}
R(t,t'):=\frac{1}{2 \eta J^2}\sum_{k,l} J_{0k} J_{l0}\chi_{kl}(t,t') \cos(\te_k-\te'_l).
\end{align*}
We next split $R(t,t')$ into the contribution from $k=l$ and the rest
\begin{align*}
	R(t,t')=:& \, \tR(t,t')+r(t,t')\\
	:=&\frac{1}{2 \eta J^2}\sum_k J_{0k}J_{k0}\chi_{kk}(t,t') \cos\bte_k(t,t')+\frac{1}{2 \eta J^2}\sum_{(k,l)} J_{0k} J_{l0}\chi_{kl}(t,t') \cos(\te_k-\te'_l).
\end{align*} 
Denoting as in the main text by $\lan\dots\ran$ the combined average over all $\J$ and the old $\om_j,\te_j^{(0)}$ we find for the moments of $\tR$ 
\begin{align*}
	\lan \tR(t,t')\ran&=\frac{1}{2 \eta J^2}\sum_k \lan J_{0k}J_{k0}\ran \lan\chi_{kk}(t,t') \cos\bte_k(t,t')\ran
	=\frac{1}{2N}\sum_k \lan\chi_{kk}(t,t') \cos\bte_k(t,t')\ran=\mathcal{O}(1)\\
	\text{and}\hspace{4.5cm}&\\
	\lan \tR(t,t')\tR(s,s')\ran
	&=\frac{1}{4 \eta^2 J^4}\sum_{k,l} \ub{\lan J_{0k}J_{k0}J_{0l}J_{l0}\ran}_{\frac{J^4}{N^2}\big(\eta^2 (1+\delta_{kl})+\delta_{kl}\big)}
	\lan\chi_{kk}(t,t') \cos\bte_k(t,t')\,\chi_{ll}(s,s') \cos\bte_l(s,s')\ran\\
	&=\frac{1}{4N^2}\sum_{k,l} \lan\chi_{kk}(t,t')\cos\bte_k(t,t')\,\chi_{ll}(s,s') \cos\bte_l(s,s')\ran\\
	&\qquad\qquad+\frac{1+\eta^2}{4 \eta^2 N^2}\sum_k\lan\chi_{kk}(t,t') \cos\bte_k(t,t')\,\chi_{kk}(s,s') \cos\bte_k(s,s')\ran.
\end{align*}	
The last term is $\mathcal{O}(1/N)$ and may be neglected.  Therefore, the fluctuations
\begin{align*}
	\lan \tR&(t,t')\tR(s,s')\ran-\lan \tR(t,t')\ran\lan \tR(s,s')\ran
\end{align*}
are small and can be neglected for $N \rightarrow \infty$. Similarly, and as in the case of symmetric couplings \cite{PRE}, also the averages 
\begin{equation*}
	\lan r(t,t')\ran, \qquad \lan r(t,t')r(s,s')\ran, \qquad \text{and} \qquad  \lan \tR(t,t')r(s,s')\ran
\end{equation*}
are negligible.
As a consequence, in the limit $N\to\infty$ we may replace $R(t,t')$ by its average:
\begin{equation}\label{resR2}
	R(t,t')\to \lan \tR(t,t')\ran=\frac{1}{2N}\sum_k \lan\chi_{kk}(t,t') \cos\bte_k(t,t')\ran.
\end{equation} 


\section{}
\label{AB}
To determine $C_2(\tau)$ we need the averages $\lan\bte_1^2(t,t')\ran$ and $\lan\bte_2(t,t')\ran$. It is advantageous to start with the average over the $\xi_a$ which to this order have correlations described by $C_0(t,t')$. For $\lan\bte_1^2(t,t')\ran_\xi$ we get as in the case $\eta=1$ 
\begin{align*}
	\lan\bte_1^2(t,t')\ran_\xi&=\int_{t'}^t\!\!dz_1\!\!\int_{t'}^t\!\!dz_2\,\lan\dte_1(z_1)\dte_1(z_2)\ran_\xi
	=\int_{t'}^t\!\!dz_1\!\!\int_{t'}^t\!\!dz_2\,\lan\ph(z_1)\ph(z_2)\ran_\xi\\
	&=\int_0^\tau\!\!dz_1\!\!\int_0^\tau\!\!dz_2\,C_0(z_1-z_2)\cos\om(z_1-z_2)
	=2\int_0^\tau\!\!du\,(\tau-u)\,C_0(u)\cos\om u,
\end{align*}
where $\tau=t-t'$ was introduced. The result for $\lan\bte_2(t,t')\ran_\xi$ does depend on $\eta$:
\begin{align*}\nonumber
	\lan\bte_2(t,t')\ran_\xi&=\int_{t'}^t\!\!dz_1\, \lan\dte_2(z_1)\ran_\xi=\int_{t'}^t\!\!dz_1\!\!\int_0^{z_1}\!\!\!\!dz_2\,
	\big[\lan\psi(z_1)\ph(z_2)\ran_\xi-\eta R_0(z_1,z_2)\sin\bte_0(z_1,z_2)\big]\\
	&= -(1+\eta)\int_0^\tau\!\!dz_1\!\!\int_{-t'}^{z_1}\!\!dz_2\,C_0(z_1-z_2)\sin\om(z_1-z_2)
	\to -(1+\eta)\tau\,\int^{\infty}_0\!\!du\,C_0(u)\sin\om u.
\end{align*}
Here the limit $t,t'\to\infty$ with fixed $\tau=t-t'$ was taken in the last step. 

The initial condition $\te^{(0)}$ dropped out of these expressions and the average over $\omega$ can be obtained via
\begin{align}\nonumber
	\lan\sin\om\tau\,\sin\om u\ran_\om&=e^{-\frac{1}{2}(\tau^2+u^2)}\sinh u\tau \spa{and}\\\label{omav}
	\lan\cos\om\tau\,\cos\om u\ran_\om&=e^{-\frac{1}{2}(\tau^2+u^2)}\cosh u\tau
\end{align}
as resulting from~\eqref{Pom}. Performing finally the $u$-integrals and plugging the results into~\eqref{C2}, we arrive at~\eqref{resC2erf}.

To obtain $R_2(t,t')$, we first need an expression for $\chi_2$. From \eqref{eq00}-\eqref{eq02} we get 
\begin{align*}
	\lan\chi_2(t,t')\ran_{\xi}=&\frac{\delta \lan\te_2(t)\ran_{\xi}}{\delta h(t')}\Bigg|_{h(t)\equiv 0}
	=-(1+\eta)\,\kappa(t-t')\!\!\int_{t'}^t\!\!dz_1\!\!\int_0^{t'}\!\!\!\!dz_2\,C_0(z_1-z_2)\cos\om(z_1-z_2)\\
	&\to -(1+\eta)\kappa(\tau)\!\!\int_0^\tau\!\!\!dz_1\!\!\int_{-\infty}^0\!\!\!\!\!\!\!dz_2\,C_0(z_1-z_2)\cos\om(z_1-z_2).
\end{align*}
We multiply by $\cos\omega\tau$ and use again~\eqref{omav} to perform the $\om$-average. Using the result together with~\eqref{resC2erf} in~\eqref{res0R2} we arrive at~\eqref{resR2erf}.


\section{}
\label{AC}

This appendix provides an approximate analysis of the correlations between the random functions $\xi(t)$ and $\zeta(t)$ introduced in~\eqref{decomppsi}. From their definitions we have
\begin{align}\label{korr}
	\lan \xi(t) \,\zeta^*(t)\ran=\eta \,J\int_0^{t}\!\! dt'\,R(t,t')\,\big\lan \xi(t) \,\, e^{-i\te(t')} \big\ran.
\end{align} 
From the structure of~\eqref{LEsc1} and the distributions of $\om$ and $\xi(t)$, we infer that the combination $\{\om,\xi(t),\te(t)\}$ has the same probability as $\{-\om,\xi^*(t),-\te(t)\}$ which implies that the average on the r.h.s. of~\eqref{korr} is a real number. Nevertheless, a general determination of this correlation is difficult because $\te(t)$ depends on all previous values of $\xi(t')$. As a first step we therefore consider small coupling strength $J$ such that we can use results from the perturbation theory described in section~\ref{sec:pt}. In particular we find from Eqs.~\eqref{eq01}, \eqref{defphpsi}, and~\eqref{res0} for $h(t)\equiv 0$
\begin{equation*}
   \te_{1}(t')= \int_0^{t'}\!\! dt'' \big[ \cos\te_{0}(t'')\,\xi_2(t'')-\sin\te_{0}(t'')\,\xi_1(t'')\big]
\end{equation*}
and therefore 
\begin{align*}
	e^{-i\te(t')}\simeq \, e^{-i\te_{0}(t')}\big( 1-i J\te_{1}(t')\big)= \, e^{-i(\om t'+\te^{(0)})} \left( 1-i J \int_0^{t'}\!\! dt'' \big[\cos\te_{0}(t'')\,\xi_2(t'')-\sin\te_{0}(t'')\,\xi_1(t'')\big] \right).
\end{align*}
To leading order in $J$ this gives rise to 
\begin{align*}
 \big\lan \xi(t) \, e^{-i\te(t')} \big\ran=&\, \big\lan \big( \xi_1(t)+i\xi_2(t)\big) \, e^{-i\te(t')} \big\ran\\
	\simeq &\, J\Big\lan e^{-i(\om t'+\te^{(0)})} \int_0^{t'}\!\! dt'' \big[ \cos\te_{0}(t'')\,\xi_2(t) \xi_2(t'')+i\sin\te_{0}(t'')\, \xi_1(t) \xi_1(t'')\big] \Big\ran \\
	\simeq &\, J \int_0^{t'}\!\! dt''C_0(t,t'')  \big\lan \,e^{-i\om(t'-t'') }\big\ran
	=2J \int_0^{t'}\!\! dt'' \,C_0(t,t'') \, C_0(t',t''),
\end{align*}
where~\eqref{Pom} was used. Plugging this into (\ref{korr}) leads to
\begin{equation*}
	\lan \xi(t) \,\, \zeta^*(t)\ran\simeq\, 2 \eta \,J^2\int_0^{t}\!\! dt'\, R_0(t,t')\int_0^{t'}\!\! dt'' \, C_0(t,t'') \, C_0(t',t'').
\end{equation*} 
Since both $C_0(t,t')$ and $R_0(t,t')$ are positive functions the equal time correlations $\lan\xi(t)\zeta^*(t)\ran$ are positive for $\eta > 0$ and negative for $\eta<0$. Using~\eqref{resC0} and~\eqref{resR0}, we can make this statement more quantitative:
\begin{align*}
	\big\lan \xi(t) \,\zeta^*(t)\big\ran\simeq &\, \eta \, \frac{J^2}{4}  \int_0^{t}\!\! dt'\,e^{-\frac{1}{2}(t-t')^2}\,
    \int_0^{t'}\!\! dt''\, e^{-\frac{1}{2}(t-t'')^2} \,e^{-\frac{1}{2}(t'-t'')^2}\\
	=&\, \eta \, J^2  \, \frac{\sqrt{\pi}}{8} \, \int_0^{t}\!\! dt' \,e^{-\frac{3}{4}(t-t')^2}\,
        \left[ \erf\left(\frac{t+t'}{2}\right)-\erf\left(\frac{t-t'}{2}\right)\right].
\end{align*}
For the stationary regime we let $t \to\infty$ and find
\begin{align*} 
	\big\lan \xi(t) \,\, \zeta^*(t)\big\ran\simeq \, \eta \, J^2 \, \frac{\sqrt{\pi}}{8} \, \int_0^{\infty}\!\! d\tau \,e^{-\frac{3}{4}\tau^2}\ \left[1-\erf\left(\frac{\tau}{2}\right)\right]
	=\, \eta \, J^2 \, \frac{\sqrt{3} \pi}{36} \simeq \, 0.15\, \eta \, J^2. 
\end{align*}
\end{widetext}

\bibliographystyle{apsrev4-2} 
\bibliography{paper} 

\end{document}